\documentclass{amsproc}

\theoremstyle{definition}

\theoremstyle{remark}

\numberwithin{equation}{section}

\begin{document}
          \title{Space Searches with a Quantum Robot}
       \author{Paul Benioff}
       \address{Physics Division, Argonne National Laboratory,
       Argonne Illinois, 60439}
       \email{pbenioff@anl.gov}
       \thanks{This work is supported by the U.S. Department of
       Energy, Nuclear Physics Division, under contract
        W-31-109-ENG-38.}

       \date{}

          \begin{abstract}
Quantum robots are described as mobile quantum computers and
ancillary systems that move in and interact with arbitrary
environments. Their dynamics is given as tasks which consist of
sequences of alternating computation and action phases. A task
example is considered in which a quantum robot searches a space
region to find the location of a system. The possibility that the
search can be more efficient than a classical search is examined
by considering use of Grover's Algorithm to process the search
results. For reversible searches this is problematic for two
reasons. One is the removal of entanglements generated by the
search process.  The other is that even if the entanglement
problem can be avoided,  the search process in 2 dimensional
space regions is no more efficient than a classical search.
However quantum searches of higher dimensional space regions are
more efficient than classical searches. Reasons why quantum
robots are interesting independent of these results are briefly
summarized.
\end{abstract}

\maketitle
\markboth{PAUL BENIOFF}{SPACE SEARCHES....}

\section{Introduction}

Quantum computers have been the subject of much study, mainly
because of computations that can be done more efficiently than on
classical computers.  Well known examples include Shor's and
Grover's algorithms, \cite{Shor,Grover}.  Quantum robots have
also been recently  described as mobile systems, with an
on board quantum computer and ancillary systems,  that move in
and interact with environments of quantum systems \cite{BenQR}.
Dynamics of quantum robots are described  as tasks consisting of
alternating computation and action phases.

Quantum robots can be used to carry out many types of tasks.
These range from simple ones such as searching a region of space
to determine the unknown location of a system to complex tasks
such as carrying out physical experiments. The fact that the
spatial searches are similar to the data base searches which are
efficiently implemented using  Grover's Algorithm \cite{Grover}
suggests that similar results might hold for use of Grover's
Algorithm to process results of a quantum search of a spatial
region. This is an example of the possible
applicability of Grover's algorithm to various physical
measurements \cite{Grover1}. If this can be done for the search
task, then one would have an example of a task that can be
carried out more efficiently by a quantum robot than by any
classical robot.

This possibility is analyzed here for a search of a 2 dimensional
space region by a quantum robot to locate a system.  It is seen
that for reversible searches with time independent unitary
dynamics, there are two problems preventing the efficient use of
Grover's Algorithm. One is that it appears impossible to remove
entanglements generated during the search process. The other
problem is that the action in the search task, which is the
equivalent of the "oracle function", which is assumed in Grover's
Algorithm to be evaluated on any argument in one step, takes many
steps to evaluate. The result is that even if the entanglement
problem is ignored, quantum searches of 2 dimensional space
regions are no more efficient than classical searches.
However quantum searches of higher dimensional space regions are
more efficient than classical searches.

The plan of this paper is to give, in the next section, a
brief description of quantum robots and a summary of how they are
different from quantum computers. This is followed, in Section
\ref{QR}, by a description of the dynamics of tasks as
sequences of alternating computation and action phases.  An
explicit description of the dynamics is given as a Feynman sum
over computation-action phase paths. An example of a quantum
robot searching a space area to determine the unknown location of
a systems is then described, Section \ref{ExQS}. The next section
is concerned with the use of Grover's algorithm to process the
search results. A very brief summary of Grover's Algorithm, in
Subsection \ref{GA}, is followed by a description of the problems
encountered, Subsection \ref{PUGA}. The paper finishes with a
discussion of why quantum robots are interesting independent of
these results.

\section{Quantum Robots}
\label{QR}
 \subsection{Comparison with Quantum Computers}
\label{CQC}

Quantum robots are similar to quantum computers in that an
important component is an on board quantum computer.  Other
systems such as a memory system $m$, an output system $o$, and a
control qubit $c$ are also present in quantum robots
\cite{BenQR}.  A relatively minor difference is that quantum
robots are mobile whereas quantum computers are stationary
relative to the environment. For quantum Turing machine models of
quantum computers the head moves but the quantum registers are
stationary. For networks of quantum gates the qubit systems move
but the gates are stationary. This is shown in physical models of
interacting qubits. Examples include ion trap models \cite{IT} or
nuclear magnetic resonance models \cite{NMR}. In these cases the
ion traps and the liquid of active molecules are stationary.

As is the case for quantum computers the effects of the
environment on the component systems of the quantum robot need to
be minimized. Methods to achieve this include the possible use of
shielding or quantum error correction codes \cite{QEC}. However
other than this the dynamical properties of the environment are
completely arbitrary.

This is quite different than the case for quantum computers. To
see this assume for quantum Turing machines that the quantum
registers are the environment of the head. Similarly for gate
networks the moving qubits may be considered as the environment
of the network of quantum gates. Here the dynamics of these
systems is quite restricted in that the states of the registers
or moving qubits can change only during interaction with the head
or the gate systems. Also the types of changes that can occur are
limited to those appropriate for the specific computation being
carried out.

This shows the main difference between quantum robots and quantum
computers, namely, that for quantum computers the states of the
qubits must not change spontaneously in the absence of
interactions related to the computation.  No such dynamical
restrictions apply to the environment of quantum robots.  The
states of environmental systems may change spontaneously whether
the quantum robot is or is not interacting with them.

Another aspect relates to the requirement that the quantum robot
cannot be a multistate system with one or a very few number of
degrees of freedom,  but must include a quantum computer. One
reason is the need for computations as part of the implementation
of any task (see below). Another is that the quantum robot may
need to be able to respond to a large number of different
environmental states.  Also a large repertoire of different (task
dependent) responses to the same environmental state must be
available.  If the total number  $N$ of needed responses is large
then the only physically reasonable approach is to make the
number of degrees of freedom of the quantum robot proportional to
$\log{N}$.  This is satisfied by including a quantum computer on
board.

\subsection{Task Dynamics}
\label{TD}

The dynamics of quantum robots are described as tasks consisting
of alternating computation and action phases \cite{BenQR}. The
purpose of each computation phase is to determine what is to be
done in the next action phase. The computation may depend on the
states of $o,m$, and the local environment as input.  The goal is
to put $o$ in one of a set $\mathcal B$ of basis states each of
which specifies an action.  During the computation the quantum
robot does not move.  Interactions with the environment, if any,
are limited to local entanglement interactions of the type
occurring in the measurement process (premeasurements in the
sense of Peres \cite{Peres}).

The purpose of an action phase is to carry out the action
determined by the previous computation phase.  The action is
determined by the state of $o$ and may include local
premeasurements of the environment state. Activities during this
phase include motion of the quantum robot and local changes in
the state of the environment. The action is independent of the
state of the on board quantum computer and the $m$ system. If $o$
is in a state in $\mathcal B$, then the state does not change
during the action.

The purpose of the control qubit $c$ is to regulate which type of
phase is active. The computation [action] phase is active only if
$c$ is in state $|1\rangle ,\; [|0\rangle ]$. Thus the last step
of the computation [action] phase is the change $|1\rangle
\rightarrow |0\rangle ,\; [|0\rangle \rightarrow |1\rangle ]$.

The overall system evolution is described here using a discrete
space time  lattice.  In this case a unitary elementary step
operator $\Gamma$ gives the overall system evolution during an
elementary time step  $\Delta$. Since $\Gamma$ has, in general,
nonzero matrix elements  between  environmental degrees of
freedom and quantum robot degrees of freedom, it  describes the
evolution of the environment and the quantum robot as well as
interactions between the environment and the quantum robot.

It is useful to decompose $\Gamma$ into two terms based on the
states of the control qubit.  If $P^{c}_{0}$ and $P^{c}_{1}$ are
projection operators on the  respective control qubit states
$|0\rangle$ and $|1\rangle$, then
\begin{equation}
\Gamma =\Gamma(P^{c}_{0}+P^{c}_{1})= \Gamma_{a}+\Gamma_{c}.
\label{step}
\end{equation}
Here $\Gamma_{a}$ and $\Gamma_{c}$ are step operators for the
action and computation phases. Interactions among environmental
degrees of freedom as well as degrees of freedom of the quantum
robot other than those taking part in the task dynamics, if any,
are also included in both operators.

Some of the conditions described for the computation and action
phases are reflected in properties that the operators
$\Gamma_{a}$ and $\Gamma_{c}$ must satisfy.  In particular, if
$P^{QR}_{x},\; P^{o}_{d}$ are projection operators for finding
the quantum robot at each lattice position $x$ ($x =
x_{1},x_{2},\cdots ,x_{d}$ in d-dimensional space) and the output
system in any state $|d\rangle$ in $\mathcal B$, then one has
\begin{eqnarray}
\Gamma_{c}P^{QR}_{x}  &  =  &  P^{QR}_{x}\Gamma_{c}  \nonumber \\
\Gamma_{a}P^{o}_{d}  &  =  &  P^{o}_{d}\Gamma_{a}
\label{conds}
\end{eqnarray}

These commutation relations express the requirements that the
position of the quantum robot does not change during the
computation phase and, except for possible entanglements with
environmental states, the state of the output system is not
changed during the action phase. These entanglements would occur
if $o$ was in a linear superposition of $\mathcal B$ states each
of which resulted in different environment states during the
action.  Another property of $\Gamma_{a}$ is that it is the
identity operator on the space of states for the on board quantum
computer and memory system degrees of freedom.  Note also that
$\Gamma_{a}$ and $\Gamma_{c}$ do not commute.

If $\Psi_{0}$ is the overall system state at time $0$ then the
state at time $n\Delta$ is given by
$\Psi_{n}=(\Gamma_{a}+\Gamma_{c})^{n}\Psi_{0}$.  The amplitude for
finding the quantum robot and environment in a state
$|w^{\prime},j\rangle$ is given by
\begin{equation}
\Psi_{n}(w^{\prime},j) = \sum_{w,i} \langle w^{\prime},j|
(\Gamma_{a}+\Gamma_{c})^{n}|w,i\rangle \Psi_{0}(w,i).
\end{equation}
Here $|w\rangle ,|w^{\prime}\rangle$ denote the states of all
environmental and quantum robot systems except the control qubit
in some suitable basis, and $i,j=0,1$ refer to the states of $c$.

All the information about the dynamics of the system is given in
the matrix elements $\langle w^{\prime},j|(\Gamma_{a}+
\Gamma_{c})^{n}|w,i\rangle$. For each $w,w^{\prime},n,i,j$ the
matrix element can be expanded in a Feynman sum over phase paths
\cite{BenQR,Brandt}. One first expands
$(\Gamma_{a}+\Gamma_{c})^{n}$ as a sum of products of
$\Gamma_{a}$ and $\Gamma_{c}$:
\begin{eqnarray}
(\Gamma_{a}+\Gamma_{c})^{n} & = & \sum_{v_{1}
=a,c}\sum_{t=1}^{n}\sum_{h_{1},h_{2}, \cdots ,h_{t} =1}^{\delta
(\sum ,n)}(P^{c}_{0}+P^{c}_{1})(\Gamma_{v_{t}})^{h_{t}}
(\Gamma_{v_{t-1}})^{h_{t-1}}, \nonumber \\
& & \mbox{} \cdots
,(\Gamma_{v_{2}})^{h_{2}}(\Gamma_{v_{1}})^{h_{1}}. \label{acsum}
\end{eqnarray}

In this expansion the number of phases is given by the value of
$t$ which ranges from $t=1$ corresponding to one phase with $n$
steps to $t=n$ corresponding to n alternating phases each of $1$
step. The duration of the $\ell th$ phase is given by the value
of $h_{\ell}$ for $\ell =1,2,\cdots ,t$.  The requirement that
the total number of steps equals $n$, or $h_{1}+h_{2} +,\cdots
,+h_{t} =n$, is indicated by the upper limit $\delta (\sum ,n)$
on the $h$ sum. The alternation of phases is shown by $v$ where
$v_{m+1} = a$ (or $c$) if $v_{m}=c$ (or $a$). The factor
$P^{c}_{0}+P^{c}_{1}$ expresses the fact that the $t$th phase may
not be completed.

Expansion in a complete set of states between each of the phase
operators $(\Gamma_{v_{\ell}})^{h_{\ell}}$ gives the desired path
sum:
\begin{eqnarray}
\langle w^{\prime},j\vert (\Gamma_{a}+\Gamma_{c})^{n}\vert
w,i\rangle & = & \sum_{t=1}^{n}\sum_{p_{2},\cdots ,p_{t}}
\sum_{h_{1},h_{2},\cdots ,h_{t} =1}^{\delta(\sum ,n)}\langle
w^{\prime},j| (\Gamma_{v_{t}})^{h_{t}}| p_{t}\rangle \nonumber \\
& & \mbox{} \cdots ,\langle p_{3}| (\Gamma_{v(2)})^{h_{2}}|
p_{2}\rangle \langle p_{2}| (\Gamma_{i})^{h_{1}}|w,i\rangle
\label{pathsum}
\end{eqnarray}
Here the sum is over all paths $p$ of states of length $t+1$ with
beginning and endpoints given by the states $|w\rangle$ and
$|w^{\prime}\rangle$. That is $|p_{1}\rangle =|w\rangle
,|p_{t}\rangle =|w^{\prime}\rangle$. The states of the control
qubit have been suppressed as they correspond to the values of
$v$. Note that $v(1) = i$.

Each term in this large sum gives the amplitude for finding $t$
alternating phases in the first $n$ steps where the $\ell th$
phase begins with all systems (except for c) in state $\vert
p_{\ell}\rangle$ and ends after $h_{\ell}$ steps in state $\vert
w_{\ell +1}\rangle$. The sums express the dispersion in the
duration or number of steps in each phase ($h$ sums), in the
number of phases ($t$ sum), and in the initial and terminal
states for each phase ($p$ sums).

\section{An Example of Quantum Searching}
\label{ExQS}

Quantum robots are well suited for carrying out search tasks. As
a simple example consider a search task where a quantum robot
searches a large square area $R$ of $N\times N$ sites to locate a
system $s$.  To keep things simple $s$ is assumed to be
motionless and located at just one unknown site. The goal of the
search is to determine the location of $s$ in $R$.

The quantum robot consists of an on board quantum computer,
memory and output systems, and a control qubit. The on board
computer is assumed here to be a quantum Turing machine
consisting of a head moving on a cyclic lattice of $O(\log {N})$
qubits. $O(-)$ denotes of the order of.  The memory system also
is a cyclic lattice which is taken here to have about the same
number of qubits and to lie adjacent to the computation lattice.
A schematic representation of the quantum robot located at a
corner (the origin) of $R$ is shown in the figure.
\begin{figure}
\vspace{3in} \caption{A Schematic Model of a Quantum Robot at the
Origin of $R$.  Both the memory system lattice (m) and Quantum
Turing machine lattice $\mathcal L$ are shown with the head $h$
that moves on the lattices. The control qubit (c) and output
system (o) are also shown. The quantum robot is greatly magnified
relative to $R$ to show details.}
\end{figure}

One method of carrying out the search is to let the coordinates
$X,Y$ with $0\leq X,Y\leq N-1$ of each point of $R$ define a
search path. If the memory is initially in state
$|X,Y\rangle_{m}$ the quantum robot, starting from the location
$0,0$ moves $X$ sites in the $x$ direction, then $Y$ sites in the
$y$ direction and looks for $s$ at its location.  After recording
the presence or absence of $s$ at the site and further
processing, if any, the quantum robot returns along the path to
the origin.

A more detailed description starts with the qubits in the memory,
$m$, and computation lattice, $\mathcal L$, in the state
$|X,Y\rangle_{m}|0\rangle_{\mathcal L}$ the output system $o$ in
state $|dn\rangle_{o}$ and a computation phase active ($c$ in
state $|1\rangle_{c}$). After copying the m state onto $\mathcal
L$ to give the state $|X,Y\rangle_{m}|X,Y\rangle_{\mathcal L}$,
the computation phase checks to see if $X=0$ or $X>0$. If $X>0$
the computation phase continues by subtracting $1$ from
$|X,Y\rangle_{\mathcal L}$ to give $|X- 1,Y\rangle_{\mathcal L}$.
It ends by changing the $o$ state to $|+x\rangle_{o}$ and the $c$
state to $|0\rangle_{c}$.

The action phase consists of one step (one iteration of
$\Gamma_{a}$) in which the quantum robot moves one lattice site
in the $+x$ direction and the $c$ state is converted back to
$|1\rangle_{c}$. The process is repeated until the state with
$X=0$ is reached on $\mathcal L$. Then the above process is
repeated for $Y$ (the $o$ state now becomes $|+y\rangle_{o}$ to
denote one step motion in the $+y$ direction) until $Y=0$ is
reached in the state of $\mathcal L$.

At this point the presence or absence of $s$ at the location
$X,Y$ of the quantum robot is recorded during a computation phase
and, after further processing, if any, the quantum robot returns
along the same path. This is done by interleaving motion of the
quantum robot in the $-y$ and $-x$ directions, with corresponding
$o$ states $|-y\rangle_{o},|-x\rangle_{o}$, with adding $1$ to
the $y$, and then $x$ components of the $\mathcal L$ state with
checking if the values $Y$ and then $X$ are reached.  This is
done by stepwise comparison with the state of $m$ which remains
unchanged.

When the state of $\mathcal L$  is the same as the state of $m$
the quantum robot has returned to the starting point at the
origin of $R$. A computation phase changes the $o$ state to
$|dn\rangle_{o}$ and transfers motion to some ballast system. As
has been noted this is necessary to preserve reversibility and
the corresponding unitarity of the dynamics \cite{Ben}.

Examples of ballast motion consist of repetitions of adding $1$
to a large lattice of $M$ qubits or emitting a particle which
moves away from $R$. In the first case with a finite number
$2^{M}$ of ballast states, the quantum robot remains in the final
state of the search degrees of freedom for a finite time only
before the search process is undone. This does not occur for the
second case with an infinite number of ballast states.

\section{Grover's Algorithm and the Quantum Search}
\label{GAQS}

Before applying Grover's Algorithm to process the results of the
search,it is useful to understand what it does and how it works.
A very brief summary, that follows Grover \cite{Grover} and Chen
et al \cite{Chen}, is given next.

\subsection{Grover's Algorithm}
\label{GA}

Suppose one has a data base $B$ of $N$ elements and a function
$f$ that takes the value $0$ on all elements except one,
$\omega$, on which $f$ has value $1$. It is assumed that $\omega$
is completely unknown and that a procedure is available for
obtaining the value of $f$ on any element of the data base in 1
step. Let each $x$ in $B$ correspond to a unique length $n$
binary string and $|x\rangle_{B}$ be the corresponding n-qubit
state.

Let the initial state for the search be given by
\begin{equation}
\phi = \frac{1}{\sqrt{N}}\sum_{x\epsilon B}|x\rangle \label{DBS}
\end{equation}
where the sum is over all $N$ elements $x$ in $B$. This
corresponds to a coherent sum over all product $|0\rangle
,|1\rangle$ states of $n$ qubits in a quantum computer if
$N=2^{n}$. This state is easily constructed from the constant $0$
state $|\underline{0}\rangle =\otimes_{j=1}^{n}|0\rangle_{j}$ by
applying the operator $(1/\sqrt{2})( \sigma_{z} +\sigma_{x})$ to
each qubit. Here $\sigma_{x},\;\sigma_{z}$ are the Pauli
matrices. This is referred to as the Walsh-Hadamard
transformation $W$. Thus $\phi = W|\underline{0}\rangle$.

Define the unitary operator $Q$ by $Q=-I_{\phi}I_{\omega}$ where
$I_{\phi} = 1-2P_{\phi}$ and $I_{\omega} = 1-2P_{\omega}$. Both
$P_{\phi}$ and $P_{\omega}$ are projection operators on the
states $\phi$ and $|\omega \rangle$. Let $|\alpha \rangle =
(1/\sqrt{N- 1})\sum_{x \neq \omega}|x\rangle$ be the coherent sum
over all states $|x\rangle$ with $x$ in $B$ and different from
$\omega$. Since $|\alpha \rangle$ and $|\omega\rangle$ are
orthonormal they form a binary basis for a 2 dimensional Hilbert
space.

One can expand $\phi$ in this basis:
\begin{equation}
\phi = \sqrt{\frac{N-1}{N}}|\alpha \rangle +
\frac{1}{\sqrt{N}}|\omega\rangle.
\end{equation}
In the same basis $Q$ has the representation
\begin{equation}
Q = \left( \begin{array}{cc} \cos{\theta} & \sin{\theta} \\ -
\sin{\theta} & \cos{\theta} \end{array} \right) \label{Qdef}
\end{equation}
where $\cos{\theta} = 1-2/N$ and $\sin{\theta} = 2\sqrt{N-1}/N$.

This shows that $Q$ acting on $\phi$ corresponds to a rotation by
$\theta$, and $m$ iterations of $Q$ correspond to a rotation by
$m\theta$. So, carrying out $m$ iterations of $Q$ on $\phi$ where
$m\theta \approx \pi /2$, rotates $\phi$ from a state that is
almost orthogonal to $|\omega\rangle$ to a state that is almost
parallel to $|\omega\rangle$. Measurement of this final state
gives with high probability, the value of $\omega$.

Grover's algorithm derives its efficiency from the fact that this
rotation is achieved with $m\sim \sqrt{N}$ whereas classically
$\sim N$ steps are needed to find $\omega$ with high probability.
The iteration of $Q$ must be stopped at the right value of $m$
because additional iterations will continue to rotate $\phi$.

Efficient implementation of this algorithm on a quantum computer,
corresponds to iteration of $Q$ on each component of $\phi$. This
requires that it is possible to determine, in a small number of
steps, if $x=\omega$ or $x\neq \omega$. This is often described
in terms of an unknown or "oracle" function $f$ on $B$, where
$f(x) =0 [=1]$ if $x\neq \omega [x=\omega]$, that can be
evaluated in one step on any $x$ in $B$.  $I_{\phi}$ is
implementable in $O(n)$ steps  as $\phi = W|\underline{0}\rangle$
and $W$ is the product of $n$ single qubit operators.

Grover \cite{Grover,Grover1} first introduced the algorithm for
searching an unstructured data base of $N=2^{n}$ elements for a
single element (f has value 1 on just one element). Since then
the algorithm has been much studied under various
generalizations. These include searches for several elements (f
has value 1 on several elements) \cite{Chen,Boyer,G2}, searches
in which $N$ is arbitrary \cite{Boyer}, and searches in which the
initial amplitude distribution of the component states is
arbitrary \cite{Lidar}. It has also been shown that the algorithm
is optimally efficient \cite{Zalka}. Further development is
described in other work \cite{Farhi,Long}. However, as has been
recently emphasized \cite{Barnes}, all these searches depend on
the fact that the evolving state is and remains a coherent
superposition of components corresponding to elements of the data
base.

\subsection{Problems with the Use of Grover's Algorithm}
\label{PUGA}

The description in Section \ref{ExQS} of the quantum search was
for the initial memory state $|X,Y\rangle_{m}$.  However if the
memory system lattice is in the initial state
\begin{equation}
\psi_{m}=(1/N)\sum_{X,Y=0}^{N-1}|X,Y \rangle_{m}, \label{Initm}
\end{equation}
then the description also applies to each component state
$|X,Y\rangle$.

As was the case for the initial state $\phi$, the state
$\psi_{m}$ can be efficiently prepared from the state
$|0\rangle_{m}$ using the Walsh-Hadamard transformation.
$(1/\sqrt 2) (\sigma_{z}+\sigma_{x})$ on each qubit of $m$.
For this initial memory state all $N^{2}$ searches are carried
out coherently. Since the path lengths range from $0$ to $2N$,
the quantum robot can search all sites of $R$ and return to the
origin in $O(N\log {N})$ steps. Since this is less than the
number of steps, $O(N^{2}\log {N})$, required by a classical
robot, the question arises if Grover's algorithm can be used to
process the final memory state to determine the location of $s$.
If this is possible, the overall search and processing should
require $O(N\log {N})$ steps which is less than that required by
a classical robot.

It is worth a digression at this point to see that Grover's
Algorithm is not applicable to the usual method of recording the
presence or absence of $s$ at a site.  To see this assume  that
$s$ is at site $X_{0},Y_{0}$ and that an extra
qubit, $r$ of the memory is set aside to record the presence or
absence of $s$.  If $r$ is  initially in state $|0\rangle_{r}$
and is changed to state $|1\rangle_{r}$ only in the presence of
$s$, then the initial memory state
$\phi_{I}=(1/N)\sum_{X,Y=0}^{N-1} |X,Y \rangle_{m}|0\rangle_{r}$
is changed to the final memory state
\begin{equation}
\phi_{f}= (1/N)(\sum_{X,Y\neq
X_{0}Y_{0}}|X,Y\rangle_{m}|0\rangle_{r} + |X_{0},Y_{0}\rangle_{m}
|1\rangle_{r}). \label{finpsi1}
\end{equation}
after the quantum robot has returned to the origin of $R$.

The idea then would be to use Grover's algorithm \cite{Grover} by
carrying out $N$ iterations of a unitary operator $U$ to amplify
the component state $|X_{0},Y_{0}\rangle_{m} |1\rangle_{r}$ at
the expense of the other components.  Following Grover and others
\cite{Grover,Chen}, define $U$ by $U=-I_{\phi_{i}}
I_{|1\rangle_{r}}$ where $I_{\phi_{i}}=1-2P_{\phi_{i}}$ and
$I_{|1\rangle_{r}}=1-2P_{|1\rangle_{r}}$.  Here $P_{\phi_{i}}$
and $P_{|1\rangle_{r}}$ are projection operators on the memory
state $\phi_{i}$ and the record state $|1\rangle_{r}$.
$P_{|1\rangle_{r}}$ is the identity on other memory degrees of
freedom.

It is clear that for this case iterations of $U$ can be carried
out efficiently. However, the problem is that the initial state
$\phi_{i}$ contains a component, $|X_{0},Y_{0}\rangle_{m}
|0\rangle_{r}$, not present in the final state $\phi_{f}$,  Eq.
\ref{finpsi1}. Also the state $\phi_{i}$ does not contain the
state $|X_{0},Y_{0}\rangle_{m}|1\rangle_{r}$.  In this case $U$
does not have a two dimensional representation in the basis pair
\begin{equation}
\frac{1}{\sqrt{N^{2}-1}}\sum_{X,Y \neq
X_{0}Y_{0}}|X,Y\rangle_{m}|0\rangle_{r}
\; ; \; \; \; |X_{0}Y_{0}\rangle_{m}|1\rangle_{r}
\end{equation}
 obtained from Eq. \ref{finpsi1}. As a result
$U$ cannot be represented as a rotation that, under iteration,
rotates the desired component to be almost parallel to the
initial state.

This problem can be avoided by changing the method of recording
the presence or absence of $s$ and not using an extra qubit $r$.
Here, following Grover \cite{Grover,Grover1},  the sign of the
component corresponding to the location of $s$ is changed.  In
this case the initial memory state, Eq. \ref{Initm}, becomes the
final memory state
\begin{equation}
\psi_{f}= (1/N)(\sum_{X,Y\neq X_{0}Y_{0}}|X,Y\rangle_{m} -
|X_{0},Y_{0}\rangle_{m}). \label{finpsi2}
\end{equation}
after return of the quantum robot. In this case $U$ is defined as
$U=-I_{\psi_{m}}I_{X_{0}Y_{0}}$ where $I_{\psi_{m}}=1-2P_{m}$ and
$I_{X_{0}Y_{0}}=1-2P_{X_{0}Y_{0}}$.  Here $P_{m}$ and
$P_{X_{0}Y_{0}}$ are projection operators on the memory states
$\psi_{m}$ and $|X_{0}Y_{0}\rangle_{m}$.

The problem here is that the only way to determine
the action of $I_{X_{0}Y_{0}}$ is by repeating the search
part of the process. This is not efficient as it requires
$O(N\log {N})$ steps. In the language of much of the work on
Grover's algorithm this corresponds to the fact that it requires
$O(N\log {N})$ steps to determine the value of the oracle
function instead of just one step as is usually assumed. In this
case the advantage of quantum over classical searching is lost
for 2 dimensional regions as use of Grover's Algorithm would
require $O(N)$ searches each requiring $O(N\log{N})$ steps.

This suggests that a method be considered in which the Grover
iterations are done prior to return when the quantum robot is at
the path endpoint.   At this point the component memory states
are entangled with the quantum robot position states as the
overall state has the form $(1/N)\sum_{X,Y}|X,Y\rangle_{m}
|X,Y\rangle_{QR}$ where $|X,Y\rangle_{QR}$ is the quantum robot
position state for the site $X,Y$.

In this case $I_{X_{0}Y_{0}}$ can be efficiently carried out on
each initial component memory state $|X,Y\rangle$ by a local
observation to see if $s$ is or is not at the site $X,Y$.  Also
the action of  $U = -I_{\psi_{m}} I_{X_{0}Y_{0}}$ on each
component memory state is given by
\begin{equation}
U|X,Y\rangle_{m}  =  -I_{m}|X,Y\rangle_{m} =
\frac{2}{\sqrt{N}}\psi_{m}-|X,Y\rangle_{m}
\end{equation}
if $|X,Y\rangle_{m}\neq |X_{0},Y_{0}\rangle_{m}$ and
\begin{equation}
U|X_{0},Y_{0}\rangle_{m}  =  I_{m}|X_{0},Y_{0}\rangle_{m} = -
\frac{2}{\sqrt{N}}\psi_{m}+ |X_{0},Y_{0}\rangle_{m}
\end{equation}
if $|X,Y\rangle_{m} = |X_{0},Y_{0}\rangle_{m}$.
Here, as before, $\psi_{m}=(1/N)\sum_{X,Y=0}^{N-1}|X,Y
\rangle_{m}$.

Here the problem is that there is no efficient way to carry out
more than one iteration of $U$.  As noted above the first
iteration can be done efficiently. However additional iterations
require that the action of $I_{X_{0}Y_{0}}$ on memory
component states $|X^{\prime},Y^{\prime}\rangle$ be evaluated for
arbitrary values of $X^{\prime},Y^{\prime}$ while the quantum
robot remains at site $X,Y$. This cannot be done efficiently as
the quantum robot has no way of knowing whether $s$ is or is not
at these different locations.   To know this  the quantum robot
must go to the site $X^{\prime},Y^{\prime}$ to see if $s$ is
there.  This is inefficient as such a trip requires $O(N\log
{N})$ steps.(Actions are efficient if they require $O(\log {N})$
steps or less. Low powers of $\log{N}$ are also acceptable.)

One sees from this that implementation of Grover's Algorithm
using either of these methods requires $O(N)$ iterations of $U$
(as $R$ has $N^{2}$ sites) where each iteration requires
$O(N\log{N})$ steps. The resulting number of steps required,
$O(N^{2}\log{N})$, is the same as that needed by a classical
robot. So quantum searches of 2 dimensional space regions
combined with Grover's Algorithm are no more efficient than
classical searches.

It is of interest to note that quantum searches of higher
dimensional space regions combined with Grover's Algorithm are
more efficient than classical searches. To see this assume a
search of a d dimensional cube of $N^{d}$ sites with the memory
in the initial state
\begin{equation}
\Psi =\frac{1}{N^{d/2}} \sum_{X_{1},\cdots
X_{d}=0}^{N-1}|X_{1},X_{2},\cdots X_{d}\rangle_{m}
\end{equation}

Carrying out Grover's Algorithm requires $O(N^{d/2})$ iterations
of $U$ where, as before,  each iteration of $U$ requires $O(N\log
{N})$ steps. This follows from the fact that the number of
dimensions appears as a multiplicative factor for the number of
steps. Also $O(dN\log{N})=O(N\log{N})$. So the overall process
requires $O(N^{(d/2)+1}\log {N})$ steps. For $d>2$ this is more
efficient than a classical search requiring $O(N^{d}\log {N})$
steps.

The discussion so far has ignored the entanglement problems.
These problems, which apply to all the above cases, result from
the fact that the task evolution, starting from the initial
unentangled product state $\psi_{m}|0\rangle_{\mathcal
L}|0,0\rangle_{QR}\cdots$, generates entanglements between the
position states $|X^{\prime},Y^{\prime}\rangle_{QR}$ of the
quantum robot and the components $|X,Y\rangle_{m}$ of the memory
state $\psi_{m}$.  In order for Grover's Algorithm to work it is
necessary to remove this entanglement at the end of each search
cycle or iteration of $U$ so that the final memory state is
$\psi_{f}$\footnote{The entanglement referred to here is
different from that  of the memory state qubits. The latter is
generated during iteration of the Grover operator and is
necessary for successful operation of Grover's Algorithm on
multiqubit states \cite{Lloyd1}.}.

This entanglement occurs because the unitary dynamics is
reversible and the number of steps needed to complete the search
task is different for different component states of $\psi_{m}$.
Here the number ranges from $O(1)$ for the path $|0,0\rangle_{m}$
to $O(2N)$ for the path $|N-1,N- 1\rangle_{m}$. This means that
the various components of the quantum robot complete an iteration
of $U$ at different times. This is independent of whether the
Algorithm is completed after or prior to return.

Because of the reversibility each component cannot simply stop
and wait until the longest search component is complete. It must
instead embark on motion of irrelevant or ballistic degrees of
freedom. This means that the memory components $|X,Y\rangle_{m}$
exchange entanglement with the quantum robot position states for
entanglement with states of ballistic degrees of freedom. Since
use of Grover's Algorithm requires the removal of this
entanglement \cite{Barnes}, the question arises whether it is
possible to insert delays into each of the memory components that
are computed, for example, after the quantum robot returns to the
origin of $R$ at the end of each cycle.  If this works then there
would be some time or step number at which the entanglement is
removed and the original product structure of the initial state
recovered, with $\psi_{f}$ replacing $\psi_{m}$.

This use of delays to remove the entanglements reversibly
requires that no memory of the magnitude of the delay be left in
the delay degrees of freedom. Otherwise one ends up with
entanglement with the delay degrees of freedom.  Also determining
the magnitude of each delay is not trivial as it depends not only
on the lengths of each of the paths but on the number of steps in
the computation phases used to determine motion along the paths.
This includes the dependence of the number of steps required to
subtract $1$ from a number $M$ on the value of $M$ (through the
number of "carry $1$" operations needed \cite{BenTBH}).

Based on these considerations it is seems doubtful that one can
use Grover's Algorithm to efficiently process the results of a
quantum search of a space region $R$.  Even if the entanglement
problem were solvable, the above results show that, for 2
dimensional space regions, use of Grover's Algorithm is no more
efficient than a classical search. For higher dimensional
searches the Algorithm is more efficient.  Note that this
conclusion is independent of the details of the quantum robot. It
applies to any quantum system such as a mobile head that contains
sufficient information on board to tell it where to go, what to
do on arrival at the endpoint, and how to return to the origin.

\section{Discussion}
\label{Disc}

In spite of these pessimistic results, quantum robots are
interesting objects of study. For instance they may be useful
test beds for study of control of quantum systems
\cite{Lloyd,Rabitz} as the dependence of the task dynamics on the
local environmental state is, for some tasks, similar to a
feedback loop.

Quantum robots and the associated task dynamics also make
clear what is and is not being done in any task. This is shown by
the quantum search task in that the quantum robot does no
monitoring or control of its behavior. It (or the on board
quantum computer) has no knowledge of where it is in $R$ at any
point or even if it is in $R$. For each component memory state
$|X,Y\rangle_{m}$ there are $X$ computation phases with the
output system o in state $|+x\rangle_{o}$ and $Y$ phases with o
in state $|+y\rangle_{o}$. These phases are interspersed with $X$
and $Y$ action phases during which anything can happen. For
example the quantum robot might move outside $R$ or it might not
move at all.  Of course for these cases it is unlikely that the
quantum robot would return to the origin at the end of the task.

This illustrates a valuable aspect of the description of the task
dynamics of quantum robots as sequences of alternating
computation and action phases. This is that, for the search task
examples described here,  it makes very clear the lack of
awareness and control the quantum robot has over what has
happened in the action phases and what it is doing. This argument
applies to the computation phases also.  For example the
"subtract $1$" steps could carry out an arbitrary change to the
memory state and the task would continue. In this case the task
would no longer be a search task but would be something else.

These considerations are also part of foundational reasons why
quantum robots and quantum computers are interesting.  If quantum
mechanics (or some extension such as quantum field theory)  is
assumed to be universally applicable, then all systems involved
in the validation of quantum mechanics are quantum systems. This
includes the systems that make theoretical computations (which
includes quantum computers) and the systems that carry out
experiments (which includes quantum robots). Thus, in some sense
quantum mechanics must describe its own validation, to the
maximum extent possible. Exploration of this and the questions of
self consistency and possible incompleteness that may occur make
this an interesting path of inquiry.

In addition quantum robots, and to some extent quantum computers,
are natural systems for investigating several questions.  In
particular what physical properties must a quantum system have
such that
\begin{itemize}
\item It is aware of its environment?
\item It has significant characteristics of intelligence?
\item It changes states of some quantum systems so that the new
states can be interpreted as text having meaning to the system
generating the text \cite{BenQVTM}?
\end{itemize}

In addition there is a sense in which the existence problem for
quantum systems having all these properties is already solved.
That is, these systems include the readers, and hopefully the
author, of this paper.

     {\footnotesize PHYSICS DIVISION, ARGONNE NATIONAL
LABORATORY, ARGONNE, IL 60439 \\
\indent {\it Current address:} Physics Division, Argonne National
Laboratory  Argonne, IL 60439 \\
\indent {\it E-mail address:} pbenioff@anl.gov}

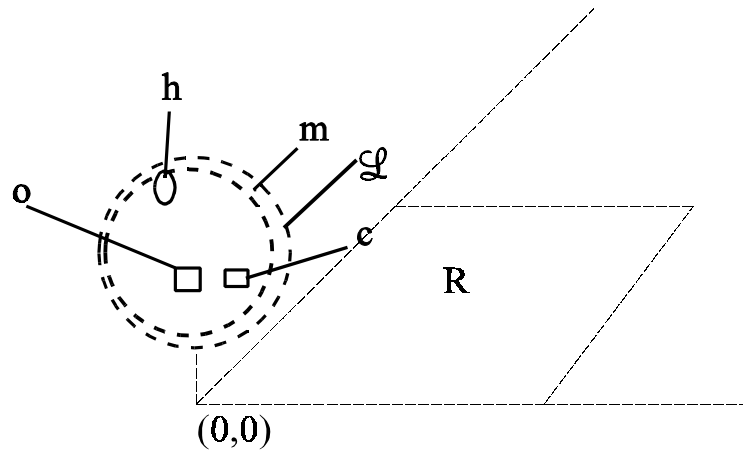
\end{document}